# Применение нейронных сетей для анализа состояния тонких пленок органических соединений на поверхности воды

Филимонов А.В., Валькова Л.А., Ососков Г.А.

### Аннотация
В данной работе нейронные сети (НС) используются как метод для построения модели процесса сжатия слоев органических соединений на поверхности раздела фаз вода - воздух. Рассматривается возможность применения НС для анализа таких слоев и интерполяции экспериментальных данных.

### Постановка задачи

В [1] изучено поведение тонких пленок 3 – нитро – 5 – трет – бутил – фталоцианина меди (CuPc*) на поверхности воды при сжатии. Данное соединение (рис. 1) исследуется для создания сенсорных наноматериалов.

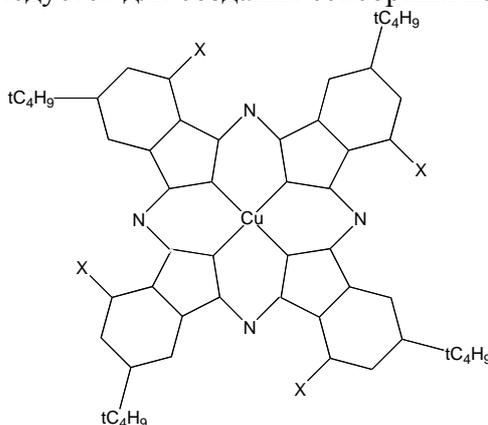

Рисунок 1. CuPc* (X=NO$_2$).

В процессе сжатия оцениваются такие физические параметры слоя как: площадь A, приходящаяся на одну молекулу, и поверхностное давление π. На основе полученных данных строятся графики: π=f(A$^{-1}$) и πA=g(π). Нас интересуют линейные участки графиков, т.к. они соответствуют определенным фазовым состояниям слоя. Используя линейные зависимости:

$$\pi = a + EA^{-1} \qquad (1)$$
$$\pi A = C + \pi A_{mol} \qquad (2)$$

можно оценить величину, характеризующую фактор сжатия слоя в определенном фазовом состоянии, E и предельную площадь, приходящуюся на одну молекулу, A$_{mol}$.

Для оценки параметров E и A$_{mol}$ при изменении условий формирования слоя (в частности, начальной поверхностной концентрации) требуется проведение нового эксперимента и расчетов. В данной работе с этой целью предлагается использовать математическую модель, построенную на основе использования нейронных сетей [2, 3].

При построении модели было показано, что нейронные сети нельзя использовать "в лоб" для моделирования экспериментальных зависимостей π=f(A$^{-1}$) и πA=g(π) при различных начальных концентрациях. Проблема связана с наличием областей сосуществования фаз (переходных точек между линейными участками). На основе этого факта был предложен метод выявления области фазовых переходов с помощью нейронных сетей.

Т.к. экспериментальные зависимости $\pi=f(A^{-1})$ и $\pi A=g(\pi)$ не удается корректно смоделировать, то во второй части данной работы предлагается альтернативный подход для оценки параметров $E$ и $A_{mol}$.

Для выбора оптимальной структуры сети использовался метод динамической оптимизации структуры сетей [4].

### Выявление областей фазовых переходов

Для выявления областей фазовых переходов используется способность нейронных сетей выявлять аномальные и противоречивые данные.

Нейронная сеть обучается на аппроксимацию зависимостей $\pi=f(A^{-1})$ и $\pi A=g(\pi)$. Если имеются противоречивые или аномальные данные, то при тестировании сети они дадут максимальную погрешность распознавания. Точки областей фазовых переходов и являются аномальными точками. Действительно, сеть не может аппроксимировать кусочные функции, а анализируемые графики зависимостей являются такими функциями, поскольку состоят из линейных и нелинейных участков. Сеть будет пытаться представить график в сглаженном виде. Если наложить друг на друга графики, смоделированные сетью и получение по экспериментальным данным, можно убедиться, что наибольшие расхождения (погрешности) приходятся на точки областей фазовых переходов (рис. 2).

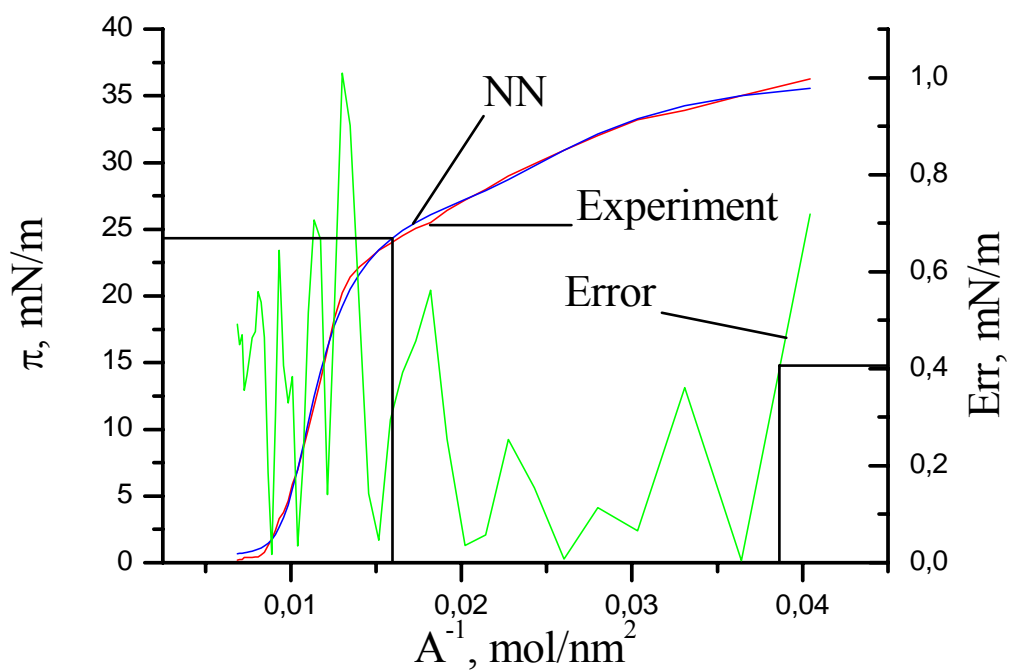

Рисунок 2. Графики зависимостей $\pi=f(A^{-1})$, построенные по экспериментальным и смоделированным сетью данным, и график зависимости величины погрешности от величины $1/A$ (начальная поверхностная концентрация $N_0 = 6{,}1 \cdot 10^7$, моль·м$^{-2}$).

Для того, чтобы определить, какие именно пики на рисунке являются фазовыми переходами необходим критерий отбора. В качестве такого критерия предложено использовать пороговое значение погрешности распознавания. Если значение погрешности распознавания больше порогового значения, то данная точка принадлежит к области фазового перехода. Порог определяется эмпирически исходя из следующих соображений. Во время эксперимента замеры площади A делаются через одинаковые интервалы. При таких условиях число экспериментальных точек, принадлежащих нелинейным участкам зависимостей $\pi=f(A^{-1})$ и $\pi A=g(\pi)$, составляет в среднем 15-20 % от общего числа точек. Таким образом, для порогового значения задается условие: выше порога должно находиться не более 15-20% экспериментальных точек.

В таблице 1 представлены результаты поиска областей фазовых переходов ($\pi_1$-$\pi_2$) и $A_{mol}$ с помощью нейросетей и без них для разных значений $N_0$ (в данном случае сравнение продемонстрировано на областях, соответствующих однофазным состояниям слоя - линейные участки на графиках $\pi A$ - $\pi$).

Таблица 1.

| $N_0 \cdot 10^7$, моль·м$^{-2}$ | С нейросетями | | Без нейросетей[1] | | Ошибка в определении $\Delta\pi_1 - \Delta\pi_2$ | Ошибка в определении $A_{mol}$ (%) |
|---|---|---|---|---|---|---|
| | $\pi_1$-$\pi_2$, мН·м$^{-1}$ | $A_{mol}$, нм$^2$ | $\pi_1$-$\pi_2$, мН·м$^{-1}$ | $A_{mol}$, нм$^2$ | | |
| 0.9 | <3 | 2.6 | <4 | 2.6 | 1 | 0 |
| 1.1 | <0.7 | 3.4 | <0.8 | 3.5 | 0.1 | 2.9 |
| | 6-19 | 1.51 | 7-20 | 1.7 | 0.1-1 | 4.4 |
| 1.7 | <9.9 | 1.63 | <10 | 1.7 | 0.1 | 4.2 |
| | >11 | 0.9 | >10 | 1.0 | 1 | 10.5 |
| 2 | <2.7 | 1.84 | <3 | 1.9 | 0.3 | 3.2 |
| | >6.9 | 0.87 | >7 | 0.9 | 0.1 | 3.4 |
| 2.1 | <3 | 2.2 | <2 | 2.1 | 1 | 4.6 |
| | 11-29 | 0.86 | 12-30 | 0.85 | 1-1 | 1.2 |
| 3 | <0.9 | 1.7 | <0.8 | 1.6 | 0.1 | 6.1 |
| | 3-7 | 1.4 | 2-6 | 1.2 | 1-1 | 15.4 |
| | 8-19 | 0.8 | 9-17 | 0.85 | 1-2 | 6.0 |
| | >37 | 0.85 | >37 | 0.85 | 0 | 0 |
| 4 | <3.4 | 1.25 | <4 | 1.3 | 0.6 | 4 |
| | 7-16 | 0.92 | 5-15 | 0.95 | 2-1 | 3.2 |
| | 27-32 | 0.94 | 28-31 | 0.95 | 1-1 | 1.0 |
| 6.1 | <3 | 1.1 | <3 | 1.1 | 0 | 0 |
| | 4-8 | 0.82 | 3-9 | 0.85 | 1-1 | 3.6 |
| | 25-35 | 0.83 | 26-33 | 0.85 | 1-2 | 2.4 |
| 8.8 | <2.6 | 1.2 | <2.5 | 1.1 | 0.1 | 8.7 |
| | 2.9-12 | 0.87 | 3-13 | 0.9 | 0.1-1 | 3.4 |
| | >13 | 0.8 | >13 | 0.8 | 0 | 0 |

[1] Используются табличные данные из [1].

Как следует из таблицы среднее расхождение значений $A_{mol}$, полученных с помощью нейросетей, и непосредственно при анализе по графикам зависимости $\pi A - \pi$, построенных по экспериментальным данным, проведенном с помощью метода наименьших квадратов составляет 4 %. Среднее расхождение значений E составляет 10.85 %. Таким образом, несмотря на некоторое расхождение в определении областей фазовых переходов, что определяется, по-видимому, несовершенством критерия отбора пороговых точек, подход с использованием нейросетей позволяет получить хорошие результаты оценки основных параметров слоя.

### Интерполяция данных

В данной работе предлагается представить E и $A_{mol}$ как функции от начальной поверхностной концентрации молекул в слое, величины обратной к площади, приходящейся на одну молекулу, и поверхностного давления.

$$E = E(N_0, A^{-1}, \pi) \qquad (4)$$
$$A_{mol} = A(N_0, \pi A, \pi) \qquad (5)$$

Одновременно с задачей нахождения E и $A_{mol}$ необходимо определить те значения $\pi$, которые соответствуют линейным участкам на графиках зависимостей $\pi = f(A^{-1})$ и $\pi A = g(\pi)$, поскольку E и $A_{mol}$ определены только для этих участков.

Предлагается моделировать зависимость $\pi = \pi(A, N_0)$. При этом имитируются условия эксперимента, когда величина A изменяется равномерно. Используя две дополнительные нейросети, обученные на моделирование зависимостей 4 и 5.

При фиксированном значении начальной концентрации на вход сети поочередно подаются смоделированные значения $\pi$ и A. На выходе сети, в зависимости от того на что эта сеть была обучена, получаются E или $A_{mol}$. Если построить графики зависимости $E = E(A^{-1}, \pi)$ и $A_{mol} = A(\pi A, \pi)$, то те значения $\pi$, которые соответствуют линейным участкам на графиках зависимостей $\pi = f(A^{-1})$ и $\pi A = g(\pi)$, образуют горизонтальные плато.

Для того, чтобы определить $A_{mol}$ и E, а также диапазон изменения $\pi$ для каждого линейного участка, достаточно прогистограммировать $A_{mol}$ и E по всем точкам. На точки областей фазовых переходов, которым соответствуют нелинейные участки зависимостей $\pi = f(A^{-1})$ и $\pi A = f(\pi)$, в среднем приходится менее 15 % экспериментальных точек. Поэтому при гистограммировании бины с наибольшим количеством попавших внутрь этого бина точек соответствуют линейным участкам (рисунок 3). Границы бина определяют начало и конец участка. Середина бина соответствует значению $A_{mol}$ и E в данном фазовом состоянии. Гистограмма, построенная для $N_0 = 0.9 \cdot 10^7$ Моль·м$^{-2}$ (рисунок 3) демонстрирует наличие одного линейного участка на кривой зависимости $\pi A = f(\pi)$, что соответствует фазовому состоянию слоя с $A_{mol} = 2.6$ нм$^2$.

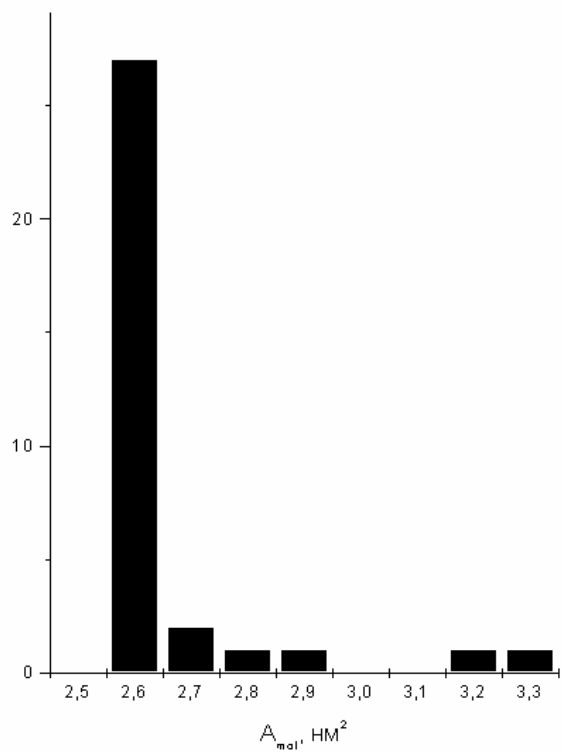

Рисунок 3. Гистограммирование величины $A_{mol}$ при начальной поверхностной концентрации $N_0=0.9 \cdot 10^7$ моль·м$^{-2}$.

В таблицах 2 и 3 сравниваются величины $A_{mol}$ и E, полученные с помощью интерполяции, с результатами, приведенными в работе [1].

Таблица 2. Интерполяция $A_{mol}$.

| $N_0 \cdot 10^7$, Моль·м$^{-2}$ | Интерполяция | | Данные из работы [1] | |
|---|---|---|---|---|
| | $\pi 1 - \pi 2$, мН·м$^{-1}$ | $A_{mol}$, нм$^2$ | $\pi 1 - \pi 2$, мН·м$^{-1}$ | $A_{mol}$, нм$^2$ |
| 1.7 | <3 | 1.84 | <10 | 1.7 |
| | >12 | 0.91 | >10 | 1.0 |
| 2.1 | <2 | 1.85 | <2 | 2.1 |
| | 14-19 | 0.9 | 12-30 | 0.85 |
| 2.2 | <1 | 1.83 | <2 | 1.6 |
| | >14 | 0.89 | >3 | 1.2 |
| 4 | <4 | 1.22 | <4 | 1.3 |
| | >6 | 0.9 | 5-15 | 0.95 |
| | Нет | | 28-31 | |

Используемый тип НС: персептроны.
Метод обучения: Resilient Propagation.
Конфигурация сети: 2 нейрона в входном слое, 17 нейронов в первом скрытом слое, 9 – во втором скрытом слое, 1 – в выходном слое.
Опорные точки: $N_0$ = 0,9; 1,1; 2; 3; 6.1; 8.8 $\cdot 10^7$ Моль·м$^{-2}$.

В среднем, расхождение между результатами, полученными экспериментальным путем и с помощью интерполяции, составляет 11,4 %, т.е. нейронная сеть интерполирует значение Amol почти с 90% точностью. Это вполне приемлемый результат. Таким образом, нейронная сеть способна аппроксимировать зависимость 5 при любых значениях начальной поверхностной концентрации, как для малых, так и для средних значений поверхностного давления. Исключение составляют большие давления ($\pi > 25$ мН/м).

Таблица 3. Интерполяция E.

| $N_0 \cdot 10^7$, Моль·м$^{-2}$ | Интерполяция | | Данные из работы [1] | |
|---|---|---|---|---|
| | $\pi 1 - \pi 2$, мН·м$^{-1}$ | $E \cdot 10^{21}$, Дж | $\pi 1 - \pi 2$, мН·м$^{-1}$ | $E \cdot 10^{21}$, Дж |
| 1.7 | <0.8 | 2.2 | <2 | 2.9 |
| 2 | <0.85 | 1.5 | <1 | 1.1 |
| 2.1 | <1.34 | 2.34 | <1 | 2 |
| 4 | <0.64 | 1.7 | <0.4 | 1 |

Используемый тип НС: персептроны.
Метод обучения: Resilient Propagation.
Конфигурация сети: 2 нейрона во входном слое, 14 нейронов в первом скрытом слое, 10 – во втором скрытом слое, 1 – в выходном слое.
Опорные точки: $N_0 = 0{,}9; 1{,}1; 3; 6.1 \cdot 10^7$ моль·м$^{-2}$.

В среднем, расхождение между результатами, полученными экспериментальным путем и с помощью интерполяции, составляет 30 %, т.е. нейронная сеть интерполирует значение E с 70 % точностью. Это означает, что при интерполяции E следует учитывать какие – то иные входные параметры.

## Обсуждение результатов

С помощью нейросетей удалось провести интерполяцию величины $A_{mol}$. Показано, что $A_{mol}$ корректно интерполируется при любых малых и средних значениях поверхностного давления, площади, приходящейся на одну молекулу, и начальной поверхностной концентрации, т.е. зависимость 5 справедлива при любых допустимых $N_0$ и $\pi$ до 25 мН/м.

Величину E удалось интерполировать только при малых значениях $\pi$ (до 5 мН/м). Тестирование показало, что сеть работает с 70 % эффективностью. Причина видимо в том, что при больших значениях поверхностного давления происходит изменение межмолекулярных взаимодействий, сопровождающееся резким уменьшением сжимаемости слоя (от 20 до 50 раз), что не учитывается в данной модели.

## Заключение

Показано, что использование нейронных сетей для построения модели процесса сжатия слоев органических соединений на поверхности раздела фаз вода – воздух позволяет проводить интерполяцию экспериментальных данных - получать предсказания параметров слоев при новых начальных условиях формирования.

Предельную площадь, приходящуюся на одну молекулу, можно представить как функцию от начальной поверхностной концентрации, площади, приходящейся на одну молекулу, и поверхностного давления. Это справедливо как для низких, так и для средних (до 25 мН/м) поверхностных давлений. Величину, характеризующую сжимаемость слоя, удалось интерполировать только при малых значениях поверхностного давления.

## Благодарности

## Литература

1. L. Valkova, N. Borovkov, M. Pisani, and F. Rustichelli, Structure of Monolayers of Copper Tetra-(3-nitro-5-tert-butyl)-Phthalocyanine at the Air-Water Interface *Langmuir*, 2001. V. 17 (12). P.3639-3642
2. Hecht-Nielsen R.,Neurocomputing, Addison-Wesley publishing company, 1989.
3. Khanna T. Foundations of neural networks, Addison-Wesley publishing company, 1990.
4. Г.А. Ососков, А.В. Филимонов, Динамическая оптимизация структуры персептронов, Сообщение ОИЯИ, Р11-2002-274, Дубна, 2002